\documentstyle[eqsecnum,aps]{revtex}

\begin{document}
\draft
\title{Bound State Spectrum of Massive Thirring Model in Rest Frame}
\author{Makoto HIRAMOTO\footnote{e-mail: hiramoto@phys.cst.nihon-u.ac.jp} and 
Takehisa FUJITA\footnote{e-mail: fffujita@phys.cst.nihon-u.ac.jp}}

\address{Department of Physics, Faculty of Science and Technology 
Nihon University, Tokyo, Japan
}
\date{\today}
\maketitle
\begin{abstract}
The bound state spectrum of the massive Thirring model is studied in the 
framework of the canonical quantization in the rest frame. First, we quantize 
the field with the massless free fermion basis states. Then, we make a 
Bogoliubov transformation. This leads to the natural mass renormalization. The 
bound state spectrum is analytically solved by the $q\bar{q}$ Fock space. It is 
found that the spectrum has the right behaviors both for the weak and for the 
strong coupling limits after the appropriate wave function regularization. This 
regularization is quite clear and the treatment is self-consistent for the 
bound state problem compared to other regularizations. Further, we show that 
the interaction between $q\bar{q}$ bosons is always repulsive and therefore 
there is no bound state in the four fermion ($qq \bar q \bar q$) Fock space. 
This confirms that there is only one bound state in the massive Thirring model. 
\end{abstract}

\pacs{PACS numbers: 11.10.Kk, 03.70.+k}

\section{Introduction}
\label{sec:level1}

The bound state spectrum of the massive Thirring model has been studied by 
several methods. Unfortunately, several different methods give 
different results on the spectrum of the bound state. Dashen {\it et al.} 
investigated the bound state mass of the Sine-Gordon model. They applied the 
WKB method to the Sine-Gordon model and obtained the soliton-antisoliton 
bound state mass \cite{DHN75}
\begin{eqnarray}
{\mathcal{M}}_{n} &=& 2{\mathcal{M}}\sin\Biggl(\frac{n\gamma}{16}\Biggl)\qquad 
n \ = \ 1,2,\cdots < \frac{8\pi}{\gamma},
\end{eqnarray}
with
\begin{eqnarray}
\gamma \ =\  \frac{\beta^{2}}{1-\frac{\beta^{2}}{8\pi}} \ 
\ =\ \frac{8\pi}{1+2\frac{g'}{\pi}},
\end{eqnarray}
where $g'$ is the coupling constant with Schwinger's regularization in the 
massive Thirring model \cite{Sc}.
In order to examine the spectrum of Eq. (1.1), they solved the Bethe-Salpeter 
equation both in the Sine-Gordon model and in the massive Thirring model, and 
obtained the bound state mass assuming that the Bethe-Salpeter equation is 
solved with the same regularization as Schwinger's one. This result corresponds 
to Eq. (1.1) if we expand Eq. (1.1) with $\gamma$ for $n=1$. Furthermore, they 
concluded that all the bound states with $n$ less than $8\pi/\gamma$ are 
stable, although the bound state with $n\ge 3$ are imbedded in the continuum 
state of $n=1$ bosons.

On the other hand, Fujita and Ogura obtained the bound state mass of the 
massive Thirring model employing the infinite momentum frame technique 
($1/K$ method) \cite{FO}. There is only one 
bound state in this method. They presented the boson mass $\mathcal{M}$ (the 
fermion-antifermion bound state) as,
\begin{eqnarray}
\mathcal{M} &=& 2m\sin\theta,
\end{eqnarray}
where $\theta$ is between $0$ and $\pi /2$ and is determined by
\begin{eqnarray}
\frac{1}{\theta\tan\theta} &=& 
\frac{g}{\pi}\Biggl[1+\frac{1}{\sin^{2}\theta}\Big(1-\frac{g}{4\pi}\Big)\Biggl],
\end{eqnarray}
where the coupling constant $g$ is Johnson's normalization \cite{Jo}. Further, 
Fujita {\it et al.} solved the Bethe ansatz equation of the massive Thirring 
model numerically \cite{FSY,FH}. They confirmed that there is only one bound 
state. They also investigated the boson-boson scattering states in 
two-particle two-hole configurations. Further, it is found that all the 
rapidity variables are real. Therefore, they claimed that there is no string 
solution which satisfies the Bethe ansatz equations in 
two-particle two-hole states.

Recent calculations of the bound state spectrum for the massive Thirring model 
have shown that several different methods give different results on the 
normalization of the coupling constant \cite{FKT}. This comes mainly from the 
normalization ambiguity of the coupling constant due to the fermion current 
regularization in the massive Thirring model. For the massless Thirring model, 
Klaiber proved that the coupling constant has a normalization ambiguity which 
arises from the fermion current regularization \cite{Kl}. In the case of the 
massive Thirring model, it is expected that the same type of the coupling 
constant ambiguity may appear.

In this paper, we calculate the bound state spectrum of the charge zero 
sector in the massive Thirring model in the rest frame which is an opposite 
limit to the $1/K$ method. The eigenvalue equation with the $q\bar{q}$ Fock 
space is solved analytically. It is shown that the spectrum has the same 
behaviors both for the weak and for the strong coupling limits as those of the 
other methods, but there still have some differences between the present 
calculation and other calculations due to the current regularization. In the 
case of the bound state problem, the regularization procedure is highly non 
trivial, since we have to sum up some types of the diagrams nonperturbatively 
in order to obtain the bound state. Here, we propose somewhat a different way 
to proceed the regularization. We carry out a regularization at the level of 
the bound state wave function in a consistent fashion. Further, in order to 
examine the number of bound states in the massive Thirring model, we consider 
the interaction between bosons in the four-fermion Fock space 
($qq\bar{q}\bar{q}$). We check analytically that the interaction between 
$q\bar{q}$ bosons is always repulsive in the $qq\bar{q}\bar{q}$ Fock space and 
find that there is no excited bound state.

This paper is organized as follows. In the next section, we briefly explain the 
quantization of the massive Thirring model. Here, we quantize the fermion field 
with the free massless fermion basis states. In this basis states, the fermion 
mass is naturally renormalized after we make the Bogoliubov transformation. In 
Section 3, we obtain the eigenvalue equation in the $q\bar{q}$ Fock space. In 
Section 4, we present a regularization scheme of the bound state wave function. 
We also discuss some properties of the bound state spectrum of this equation. 
In section 5, we evaluate the interaction between $q\bar{q}$ bosons of the 
massive Thirring model in the four-fermion Fock space ($qq\bar{q}\bar{q}$). It 
is found that the four-fermion Fock space does not produce any bound state.
Finally, we summarize in Section 6 what we have clarified from the present 
paper.

\section{The Hamiltonian of The Massive Thirring Model}
The massive Thirring model is a (1+1) dimensional field theory which is 
described by the following Lagrangian density,
\begin{eqnarray}
{\cal{L}} = \bar{\psi}(i\gamma^{\mu}\partial_{\mu} - m_{0})\psi 
            -\frac{1}{2}gj_{\mu}j^{\mu}.
\end{eqnarray}
The fermion current $j^{\mu}$ is defined by
\begin{eqnarray}
j^{\mu} = :\bar{\psi}\gamma^{\mu}\psi:,
\end{eqnarray}
where we choose $\gamma$ matrices as 
\begin{eqnarray}
\gamma^{0}=
\left(
      \begin{array}{cc}
       0 & 1 \\
       1 & 0
      \end{array}
\right), \quad
\gamma^{1}=
\left(
      \begin{array}{cc}
       0 & -1 \\
       1 & 0
      \end{array}
\right), \quad
\gamma^{5}=
\left(
      \begin{array}{cc}
       1 & 0 \\
       0 & -1
      \end{array}
\right). \quad
\end{eqnarray}
In this representation, the Hamiltonian is given as
\begin{eqnarray}
H = \int dx 
\left[
-i\psi^{\dagger}_{1}\frac{\partial}{\partial x}\psi_{1}
+i \psi^{\dagger}_{2}\frac{\partial}{\partial x}\psi_{2}+
m_{0}(\psi^{\dagger}_{1}\psi_{2}+\psi^{\dagger}_{2}\psi_{1})
+2g\psi^{\dagger}_{1}\psi^{\dagger}_{2}\psi_{2}\psi_{1}
\right].
\end{eqnarray}
Now, we quantize the fermion field in a box $L$
\begin{eqnarray}
\psi(x)=\frac{1}{\sqrt{L}}\sum_{n}\left(
      \begin{array}{c}
       a_{n} \\
       b_{n}
      \end{array}
\right)e^{ip_{n}x},\\
\{a_{i},a_{j}^{\dagger}\}=\{b_{i},b_{j}^{\dagger}\}=\delta_{i,j}.
\end{eqnarray}
Then, the Hamiltonian becomes
\begin{eqnarray}
H&=&\sum_{n}\left[p_{n}(a_{n}^{\dagger}a_{n}-b_{n}^{\dagger}b_{n})
+m_{0}(a_{n}^{\dagger}b_{n}+b_{n}^{\dagger}a_{n})+
\frac{2g}{L}\tilde{j}_{1,p_{n}}\tilde{j}_{2,-p_{n}}\right],
\end{eqnarray}
where currents in the momentum representation $\tilde{j}_{1,p_{n}}$ and 
$\tilde{j}_{2,p_{n}}$ are given by
\begin{eqnarray}
\tilde{j}_{1,p_{n}}=\sum_{l}a_{l}^{\dagger}a_{l+n}\\
\tilde{j}_{2,p_{n}}=\sum_{l}b_{l}^{\dagger}b_{l+n}.
\end{eqnarray}
For the free field theory ($g=0$), the Hamiltonian can be exactly diagonalized 
by a Bogoliubov transformation which mixes left- and right-handed fermions. On 
the other hand, this is not possible for interacting cases. However, for the 
massive Thirring model, we can get reliable results as we consider the bound 
state spectrum. Now, we introduce new fermion operators by a Bogoliubov 
transformation,
\begin{eqnarray}
\frac{c_{n}-d_{-n}^{\dagger}}{\sqrt{2}}&=&
\cos{\frac{\theta_{n}}{2}}a_{n}+\sin{\frac{\theta_{n}}{2}}b_{n},
\\
\frac{c_{n}+d_{-n}^{\dagger}}{\sqrt{2}}&=&
-\sin{\frac{\theta_{n}}{2}}a_{n}+\cos{\frac{\theta_{n}}{2}}b_{n}.
\end{eqnarray}
First, we consider the free field theory. In this case, the free field 
Hamiltonian $H_{0}$ can be written in terms of the new operators $c_{n}$ and 
$d_{n}$
\begin{eqnarray}
H_{0} &=&
\sum_{n}\Big[\left(p_{n}\sin{\theta_{n}}
+m_{0}
\cos{\theta_{n}}\right)
(c_{n}^{\dagger}c_{n}+d_{-n}^{\dagger}d_{-n})\nonumber \\
&&
+(-p_{n}\cos{\theta_{n}}+m_{0}\sin{\theta_{n}})
(c_{n}^{\dagger}d_{-n}^{\dagger}+d_{-n}c_{n})\Big].
\end{eqnarray}
The condition that the terms proportional to $(c^{\dagger}d^{\dagger}+cd)$ must 
vanish determines the theta parameter $\theta_{n}$. This is given as
\begin{eqnarray}
\tan\theta_{n} &=& \frac{p_{n}}{m_{0}}.
\end{eqnarray}
Then, the new free field Hamiltonian $H_{0}'$ becomes
\begin{eqnarray}
H_{0}' &=& \sum_{n}\sqrt{p_{n}^2+m_{0}^2}
\ (c^{\dagger}_{n}c_{n}+d^{\dagger}_{n}d_{n}).
\end{eqnarray}

Next, we make a Bogoliubov transformation Eq. (2.10) and (2.11) for the full 
Hamiltonian Eq. (2.7). Then, the new Hamiltonian $H'$ becomes
\begin{eqnarray}
H' = \tilde{H_{0}}+H_{cd}+H_{C}+H_{A}+H_{R}+H_{non},
\end{eqnarray}
where $\tilde{H_{0}}$ denotes the free field Hamiltonian,
\begin{eqnarray}
\tilde{H_{0}} &=&
\sum_{n}\Biggl[\left\{p_{n}\sin{\theta_{n}}
+\left(m_{0}+\frac{g}{L}\mathcal{A}\right)
\cos{\theta_{n}}\right\}
(c_{n}^{\dagger}c_{n}+d_{-n}^{\dagger}d_{-n})\Biggl],
\end{eqnarray}
where $\mathcal{A}=\sum\cos\theta_{n}$ is related to the mass renormalization. 
We rewrite the fermion mass in terms of the renormalized mass $m$
\begin{eqnarray}
m &=& m_{0}+\frac{g}{L}\mathcal{A}.
\end{eqnarray}
Therefore, we can naturally renormalize the fermion mass. 

The interacting Hamiltonian can be written as
\begin{eqnarray}
H_{C} &=& -\frac{2g}{L}\sum_{k,p,q}\Big\{
\sin\frac{\theta_{p}}{2}\sin\frac{\theta_{p+k}}{2}
\sin\frac{\theta_{q}}{2}\sin\frac{\theta_{q-k}}{2}
\ c_{p}^{\dagger}c_{p+k}d_{-q+k}^{\dagger}d_{-q}
\nonumber\\ && \qquad
+\cos\frac{\theta_{q}}{2}\cos\frac{\theta_{q-k}}{2}
\cos\frac{\theta_{p}}{2}\cos\frac{\theta_{p+k}}{2}
\ c_{q}^{\dagger}c_{q-k}d_{-p-k}^{\dagger}d_{-p}
\Big\},\\
H_{A} &=& -\frac{2g}{L}\sum_{k,p,q}\Big\{
\sin\frac{\theta_{p}}{2}\cos\frac{\theta_{q-k}}{2}
\cos\frac{\theta_{p+k}}{2}\sin\frac{\theta_{q}}{2}
\ c_{p}^{\dagger}c_{q-k}d_{-p-k}^{\dagger}d_{-q}
\nonumber\\ && \qquad
+\cos\frac{\theta_{q}}{2}\sin\frac{\theta_{p+k}}{2}
\sin\frac{\theta_{q-k}}{2}\cos\frac{\theta_{p}}{2}
\ c_{q}^{\dagger}c_{p+k}d_{-q+k}^{\dagger}d_{-p}
\Big\},\\
H_{R} &=& -\frac{2g}{L}\sum_{k,p,q}\Big\{
\sin\frac{\theta_{p}}{2}\cos\frac{\theta_{q}}{2}
\sin\frac{\theta_{p+k}}{2}\cos\frac{\theta_{q-k}}{2}
\ c_{p}^{\dagger}c_{q}^{\dagger}c_{p+k}c_{q-k}
\nonumber\\ && \qquad
+\cos\frac{\theta_{p}}{2}\sin\frac{\theta_{q}}{2}
\cos\frac{\theta_{p+k}}{2}\sin\frac{\theta_{q-k}}{2}
\ d_{-p-k}^{\dagger}d_{-q+k}^{\dagger}d_{-p}d_{-q}
\Big\},
\end{eqnarray}
and $H_{non}$ denotes the term which does not conserve fermion antifermion 
number. The interacting Hamiltonian $H_{cd}$ describes the 
$(c^{\dagger}d^{\dagger}+cd)$ terms
\begin{eqnarray}
H_{cd} &=& \sum_{n}
(-p_{n}\cos{\theta_{n}}+m\sin{\theta_{n}})
(c_{n}^{\dagger}d_{-n}^{\dagger}+d_{-n}c_{n}).
\end{eqnarray}
It is interesting to note that the Hamiltonian $H_{cd}$ has the same shape as 
the free field Hamiltonian $H_{0}$. The important point is that the free 
field Hamiltonian has the bare mass $m_{0}$ after the Bogoliubov transformation 
while we have the renormalized mass in Eq. (2.21). We impose the same condition 
as the free field Hamiltonian. Thus, we obtain
\begin{eqnarray}
\tan\theta_{n} &=& \frac{p_{n}}{m},
\end{eqnarray}
which is just the condition for the free field Hamiltonian except $m$. Then, 
$\tilde{H_{0}}$ is modified as
\begin{eqnarray}
\tilde{H_{0}} &=& \sum_{n}E_{p_{n}}(c^{\dagger}_{n}c_{n}+d^{\dagger}_{n}d_{n}),
\end{eqnarray}
where $E_{p_{n}}=\sqrt{p_{n}^2+m^2}$.

This is all that is necessary to evaluate the bound state spectrum of the 
massive Thirring model.

\section{Bound State Spectrum within $q\bar{q}$ Fock Space}
In order to obtain physical quantities, we have to diagonalize the Hamiltonian. 
In this case, we have to prepare basis states. Here, we employ the Fock space 
expansion. We limit the Fock space to $q \bar{q}$ only, since there is no 
particle creation in the massive Thirring model, and therefore there is no 
mixture between $q\bar{q}$ and $qq\bar{q}\bar{q}$ Fock spaces.

Now, the Fock space for the $q\bar{q}$ state can be written
\begin{eqnarray}
|q\bar{q}\rangle &=& \sum_{n}f_{n}c_{n}^{\dagger}d_{-n}^{\dagger}|0\rangle,
\end{eqnarray}
where $f_{n}$ is a wave function in momentum space and satisfies the 
normalization condition, $\sum |f_{n}|^2=1$. The energy eigenvalue of the 
Hamiltonian with the $q\bar{q}$ Fock space can be written as
\begin{eqnarray}
\mathcal{M} &=& 2\sum_{n}|f_{n}|^{2}E_{p_n}
-\frac{g}{L}\sum_{l,n}f_{l}^{\dagger}f_{n}
(1+\sin\theta_{l}\sin\theta_{n}+\cos\theta_{l}\cos\theta_{n}).
\end{eqnarray}
Equivalently, we can write the equation by making variations with respect to 
$f_{n}$,
\begin{eqnarray}
{\mathcal{M}}f_{n} &=& 2E_{p_n}f_{n}
-\frac{g}{L}\sum_{l}f_{l}(1+\sin\theta_{l}\sin\theta_{n}
+\cos\theta_{l}\cos\theta_{n}),
\\
&=& 2E_{p_n}f_{n}
-\frac{g}{L}\sum_{l}f_{l}\left(
1+\frac{m^2}{E_{p_{n}}E_{p_{l}}}+\frac{p_{n}p_{l}}{E_{p_{n}}E_{p_{l}}}\right).
\end{eqnarray}
Converting the sum to the integral, we finally obtain
\begin{eqnarray}
{\mathcal{M}}f(p) &=& 2E_{p}f(p)-\frac{g}{2\pi}\int dq f(q)
\left(
1+\frac{m^2}{E_{p}E_{q}}+\frac{pq}{E_{p}E_{q}}\right).
\end{eqnarray}
This equation is solved analytically, because it is a separable type 
interaction. 

Now, we solve the integral equation (3.5). Here, we assume that the wave 
function satisfies the symmetric condition $f(-p)=f(p)$. In this case, we can 
drop the last term of Eq. (3.5). We define the following quantities $A$ and $B$ 
\begin{eqnarray}
A &=& \int_{-\infty}^{\infty} dp f(p),\\
B &=& \int_{-\infty}^{\infty} dp \frac{f(p)}{E_{p}}.
\end{eqnarray}
Using $A$ and $B$, we can solve Eq. (3.5) for $f(p)$ and obtain
\begin{eqnarray}
f(p) &=& \frac{g/2\pi}{2E_{p}-\mathcal{M}}\left(A+\frac{m^2}{E_{p}}B\right).
\end{eqnarray}
Putting this $f(p)$ back into Eqs. (3.6) and (3.7), we obtain the matrix equations
\begin{eqnarray}
A &=& \frac{g}{2\pi}\int_{0}^{\Lambda} \frac{2dp}{2E_{p}-\mathcal{M}}
\left(A+\frac{m^2}{E_{p}}B\right),\\
B &=& \frac{g}{2\pi}\int_{0}^{\infty} \frac{2dp}{(2E_{p}-{\mathcal{M}})E_{p}}
\left(A+\frac{m^2}{E_{p}}B\right).
\end{eqnarray}
Now, the regularization is necessary for the first equation, since the 
integral has the divergent term $\log[\Lambda/m+\sqrt{1+(\Lambda/m)^2}]$. This 
arises from the current-current interaction at the same space-time point. This 
must be regularized. In the next section, we treat the wave function 
regularization procedure.

\section{Wave Function Regularization}
The regularization procedure is well defined for the perturbative treatment. 
There is no serious problem to renormalize divergent terms in the massive 
Thirring model. However, if we treat bound state problems, then the 
regularization procedure is highly nontrivial. This is because we have to sum 
up some of the diagrams nonperturbatively in order to obtain any bound states. 
For example, Dashen {\it et al.} \cite{DHN75} carried out the regularization of 
the coupling constant perturbatively for the bound state. There, they regularize 
the divergent term by first solving the eigenvalue equation and then by adding 
a new counter term to cancel the divergent part in the energy eigenvalue. 

Here, we propose somewhat a different way to proceed the regularization. We 
carry out the regularization at the level of wave function in a consistent 
fashion. The origin of the divergent term comes from the nature of the 
current-current interaction $j_{\mu}(x)j^{\mu}(y)$ which has the divergent at 
the same space-time point $x=y$.  

In our case, the divergent term $\log[\Lambda/m+\sqrt{1+(\Lambda/m)^2}]$ arises from 
the integration of $E_{p}^{-1}$. We subtract $gA/4\pi E_{p}$ from $f(p)$,
\begin{eqnarray}
\bar{f}(p) &=& f(p)-\frac{g}{2\pi}\frac{A}{2E_{p}}.
\end{eqnarray}
In this case, we obtain
\begin{eqnarray}
\bar{f}(p) &=& 
\frac{g/2\pi}
{(2E_{p}-{\mathcal{M}})E_{p}}\left(\frac{\mathcal{M}}{2}\bar{A}+m^{2}\bar{B}\right).
\end{eqnarray}
With this $\bar{f}(p)$, we obtain new matrix equations as
\begin{eqnarray}
\bar{A} &=& \frac{g}{2\pi}\alpha^{-1}
\left(\frac{\pi}{2}+\tan^{-1}
\frac{\mathcal{M}}{2\alpha}\right)
\left(\frac{\mathcal{M}}{2}\bar{A}+m^2\bar{B}\right),\\
\bar{B} &=&\frac{g}{2\pi}\frac{2}{\mathcal{M}}\left[\alpha^{-1}
\left(\frac{\pi}{2}+\tan^{-1}
\frac{\mathcal{M}}{2\alpha}\right)-\frac{\pi}{2m}\right]
\left(\frac{\mathcal{M}}{2}\bar{A}+m^2\bar{B}\right),
\end{eqnarray}
where, $\alpha=\sqrt{m^2-({\mathcal{M}}/2)^2}$. Here, the divergent term is 
cancelled out. Therefore, the above equations are written by the  finite 
quantities only. The divergent quantities are removed in a very nice and 
consistent fashion. We note that there are other regularizations for the 
current-current interaction at the same space-time point. One good example 
of the regularization could be $e^{ip\varepsilon}\ (\varepsilon \to 0)$. 
However, if one wants to apply $e^{ip\varepsilon}$ regularization to the 
bound state problem, one cannot regularize the wave function in a consistent 
way.

Now, we obtain the eigenvalue equation,
\begin{eqnarray}
\frac{g}{2\pi}\left[
\left(\frac{\mathcal{M}}{2\alpha}+\frac{2m^2}{\mathcal{M}\alpha}\right)
\left(\frac{\pi}{2}+\tan^{-1}\frac{\mathcal{M}}{2\alpha}\right)
-\frac{\pi m}{\mathcal{M}}\right] &=& 1.
\end{eqnarray}
Introducing a new variable $\theta$ by
\begin{eqnarray}
\mathcal{M} &=& 2m\sin\theta,
\end{eqnarray}
Eq. (4.5) becomes
\begin{eqnarray}
\frac{g}{2\pi}\left[
\left(\tan\theta+\frac{1}{\sin\theta\cos\theta}\right)
\left(\frac{\pi}{2}+\theta\right)
-\frac{\pi}{2\sin\theta}\right] &=& 1,
\end{eqnarray}
where $\theta$ is between $0$ and $\pi/2$.

Now, we check the strong coupling limit $\theta\sim 0$. This corresponds to 
${\mathcal{M}}=2m\sin\theta\sim 2m\theta$. From Eq. (4.7), we get
\begin{eqnarray}
\theta \sim \frac{2}{3\pi}\left(2-\frac{g}{\pi}\right).
\end{eqnarray}
Therefore, the bound state spectrum becomes
\begin{eqnarray}
{\mathcal{M}} \sim \frac{4}{3\pi}\left(2-\frac{g}{\pi}\right)m.
\end{eqnarray}
Namely, $\mathcal{M}$ becomes zero for $g/\pi=2$.  Beyond that, 
there is no bound state and the theory is not well defined. This is 
precisely what is predicted by Johnson \cite{Jo}. This is because  we use the 
massless free fermion basis states. In this case, the coupling constant 
normalization is bound to use Johnson's normalization as Klaiber shows 
\cite{Kl}. 
This situation can be seen easily in terms of bosonized form of 
the Thirring model. Since the free fermion current satisfies boson's 
commutation relations, the bosonized Hamiltonian of the Thirring model can be 
written as
\begin{eqnarray}
H &=& \sum_{p}\left[\left(1-\frac{g}{2\pi}\right)\Pi^{\dagger}(p)\Pi(p)+
\left(1+\frac{g}{2\pi}\right)p^2\Phi^{\dagger}(p)\Phi(p)\right],
\end{eqnarray}
where $\Phi(p)$ and $\Pi(p)$ are related to the free fermion current as
\begin{eqnarray}
\tilde{J}_{0}(p) = \tilde{j}_{1,p}+\tilde{j}_{2,p}
                 = ip\sqrt{\frac{L}{\pi}}\Phi(p)\\
\tilde{J}_{0}(p) = \tilde{j}_{1,p}-\tilde{j}_{2,p}
                 = -\sqrt{\frac{L}{\pi}}\Pi(p).
\end{eqnarray}
Therefore, we can see that
\begin{eqnarray}
-2\leq \frac{g}{\pi} \leq 2
\end{eqnarray}
which is just Johnson's constraint. For the massive Thirring model, we should 
add the mass term $m \bar \psi \psi $ to the Lagrangian. However, the mass 
term can be described only by $\Phi (p)$ without kinetic terms, and therefore 
it does not influence the condition that $g/\pi$ should be smaller than 2.

Next, we consider the weak coupling limit $(g/\pi\sim 0)$. In this case, we put 
$\theta=\pi/2-\beta\ (\beta\sim 0)$. Then, we obtain the bound state spectrum 
\begin{eqnarray}
{\mathcal{M}} = 2m\cos\beta 
&=& m\left[2-g^2+\left(2+\frac{\pi}{2}\right)\frac{g^3}{\pi}+O(g^4) \right].
\end{eqnarray}
This result can be compared with the prediction of Fujita and Ogura in the 
$1/K$ method calculation
\begin{eqnarray}
\mathcal{M} &=& 
m\left[2-g^2+\left(4+\frac{1}{4}\right)\frac{g^3}{\pi}+O(g^4) \right].
\end{eqnarray}
On the other hand, the WKB formula Eq. (1.1) with Johnson's regularization 
becomes
\begin{eqnarray}
\mathcal{M} &=& m\left[2-g^2+3\frac{g^3}{\pi}+O(g^4) \right].
\end{eqnarray}
As can be seen, the coefficients of $g^3$ are different from each other. This 
suggests that each method has nice features both for the weak and for the 
strong coupling limits, but there still have ambiguities which are related to 
the current-current regularization.

In Fig. 1, we show numerical results of Eq. (4.7), the $1/K$ method, the WKB 
method and the Bethe ansatz solution. As can be seen from Fig. 1, the 
difference between the present calculation and the WKB method is surprisingly 
small, while the difference between the present calculation and the $1/K$ 
method is not so small. This suggests that the regularization may well be 
related to the total momenta $K$ of $q\bar{q}$. 

\section{Four-Fermion Fock Space}
In this section, we consider the four-fermion Fock space. To include 
the four-fermion Fock space, we must be careful for treating transitions from 
$q\bar{q}$ to $qq\bar{q}\bar{q}$ Fock space. It is well known that the 
transition that changes fermion numbers is forbidden in the massive Thirring 
model. This is due to an infinite number of conservation laws that must be 
obeyed by the external momenta \cite{BKT,Be}. Here, we consider the $H_{R}$ 
term. This contributes to the higher order Fock space

Now, we calculate the expectation value of $H_{R}$ in the $qq\bar{q}\bar{q}$ 
Fock space. The $qq\bar{q}\bar{q}$ Fock space is defined by
\begin{eqnarray}
|qq\bar{q}\bar{q}\rangle &=& 
\sum_{p_{1}p_{2}q_{1}q_{2}}f^{(4)}(p_{1},p_{2};q_{1},q_{2})
c^{\dagger}_{p_{1}}c^{\dagger}_{p_{2}}d^{\dagger}_{q_{1}}d^{\dagger}_{q_{2}}
|0\rangle \delta_{p_{1}+p_{2}+q_{1}+q_{2},0},
\end{eqnarray}
where $f^{(4)}(p_{1},p_{2};q_{1},q_{2})$ has an antisymmetric property;
\begin{eqnarray}
f^{(4)}(p_{2},p_{1};q_{1},q_{2})=f^{(4)}(p_{1},p_{2};q_{2},q_{1})
=-f^{(4)}(p_{1},p_{2};q_{1},q_{2}).
\end{eqnarray}
Thus, we obtain
\begin{eqnarray}
\mathcal{E}_{R} &=& \langle qq\bar{q}\bar{q}|H_{R}|qq\bar{q}\bar{q}\rangle 
\nonumber\\
&=& \frac{8g}{L}\sum_{p_{1}p_{2}q_{1}q_{2}}
|f^{(4)}(p_{1},p_{2};q_{1},q_{2})|^{2}
(\sin\theta_{p_{1}}\cos\theta_{p_{2}}-\cos\theta_{p_{1}}\sin\theta_{p_{2}})^{2}.
\end{eqnarray}
Now, we  see that ${\cal E}_R$ is always positive. Therefore, this is 
repulsive and thus there is no bound state in the four fermion Fock space. From 
this calculation, we  conclude that there is only one bound state in the 
massive Thirring model. 

\section{Conclusions}
We have presented the bound state spectrum in the massive Thirring model. This 
is based on the $q\bar{q}$ Fock space in the rest frame. The eigenvalue 
equation with the $q\bar{q}$ Fock space is solved analytically. It is shown 
that the bound state spectrum has right behaviors both for the weak and for the 
strong coupling limits after the appropriate wave function regularization. This 
is indeed confirmed since the $q \bar q$ Fock space has no transition to higher 
Fock space due to an infinite number of conservation law. Further, in order to 
confirm that there is only one bound state, we carry out the analytic 
evaluation of the interaction term in the $qq\bar{q}\bar{q}$ Fock space. We 
show that the interaction between the bosons of $q \bar q$ Fock space is 
repulsive and therefore there is no bound state in the four-fermion Fock space. 

Although there are still some differences between the present result and other 
methods, we can reliably calculate the bound state spectrum after the 
appropriate wave function regularization. At the present stage, it is not so 
clear whether the regularization ambiguity of the massive Thirring model can 
be different from the massless Thirring model. It may well be that the 
regularization ambiguity is related to some hidden symmetry which is not yet 
clearly understood up to now.

\acknowledgments

We thank K. Yazaki for careful reading of the manuscript  and for stimulating 
discussions.

\section*{Figure Captions}

\begin{figure}
\caption{We show the bound state spectrum for the massive Thirring model with
the $q\bar{q}$ Fock space as the function of $g/\pi$. Here, the coupling 
constant is Johnson's normalization. The solid line is calculated by Eq. (4.7) 
while the dotted line is calculated by Eq. (1.4). The dashed line is calculated 
by Eq. (1.1) with $n=1$. The dot-dashed line indicates the twice of Eq. (1) 
with $n=1$. We also plot the Bethe ansatz results with error bars. }
\label{fig1}
\end{figure}


\begin{references}

\bibitem{BT} H. Bergknoff and H. B. Thacker, Phys. Rev. Lett. {\bf 42} 
(1979) 135.
\bibitem{Th} H. B. Thacker, Rev. Mod. Phys. {\bf 53} (1981) 253.
\bibitem{DHN75} R. F. Dashen, B. Hasslacher and A. Neveu, Phys. 
Rev. D{\bf 11} (1975) 3424.
\bibitem{FO} T. Fujita and A. Ogura, Prog. Theor. Phys. {\bf 89} (1993) 23.
\bibitem{FSY} T. Fujita, Y. Sekiguchi and K. Yamamoto, Ann. Phys. {\bf 255} 
(1997) 204.
\bibitem{FH} T. Fujita and M. Hiramoto, Phys. Rev. {\bf D58} (1998) 125019.
\bibitem{Ca} M. Cavicchi, Int. J. Mod. Phys. {\bf A10} (1995) 167.
\bibitem{Sc} J. Schwinger, Phys. Rev. Lett. {\bf 3} (1959) 296.
\bibitem{Jo} K. Johnson, Nuovo Cimento {\bf 20} (1961) 773.
\bibitem{FKT} T. Fujita, T. Kake and H. Takahashi, 
Ann. Phys. {\bf 282} (2000) 100.
\bibitem{Kl} B. Kleiber, in Lectures in Theoretical Physics, 1967, 
edited by A. Barut and W. Britten (Gordon and Breach, 1968).
\bibitem{Na} N. Nakanishi, Prog. Theor. Phys. {\bf 57} (1997) 580.
\bibitem{TO} H. Takahashi and A. Ogura, Prog. Theor. Phys. {\bf 105} 
(2001) 495.
\bibitem{BKT} B. Berg, M. Karowski, and H. -J. Thun, Phys. Lett. {\bf B62} 
(1976) 187.
\bibitem{Be} B. Berg, Nuovo Cimento {\bf 41A} (1977) 58.
\end{references}
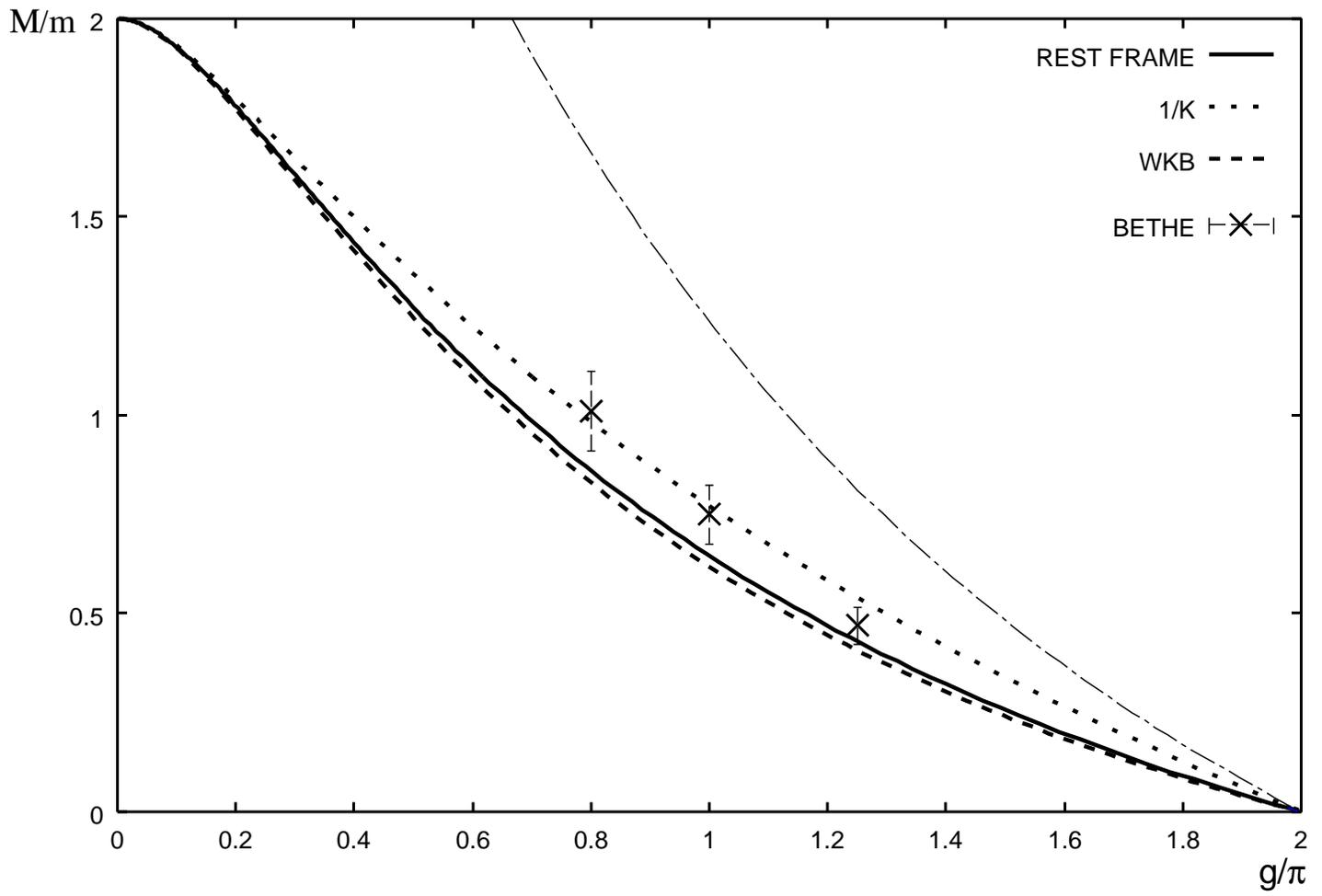
\end{document}